\documentclass[letter,twocolumn,superscriptaddress]{revtex4}

\usepackage{epsfig}
\usepackage{graphicx}

\begin{document}
\title{Polar distortions in hydrogen bonded organic ferroelectrics}

\author{Alessandro Stroppa}
\affiliation{CNR-SPIN, L'Aquila, Italy}
\email{alessandro.stroppa@spin.cnr.it}

\author{Domenico Di Sante}
\affiliation{University of L'Aquila, Physics Department,
Via Vetoio, L'Aquila}

\author{Sachio Horiuchi}
\affiliation{National Institute of Advanced Industrial Science and Technology (AIST)
Tsukuba, Ibaraki, 305-8562, Japan\\
and\\
CREST, JST, Tsukuba, 305-8562, Japan.}

\author{Yoshinori Tokura}
\affiliation{Department of Applied Physics, The University of Tokyo
Hongo Bunkyo-ku, Tokyo, 113-8656, Japan\\
and\\
Correlated Electron Research Group, RIKEN Advanced Science Institute,
Wako 351-0198, Japan}

\author{David Vanderbilt}
\affiliation{Department of Physics and Astronomy, Rutgers University,
136 Frelinghuysen Road, Piscataway, New Jersey 08854-8019, USA}

\author{Silvia Picozzi}
\affiliation{CNR-SPIN, L'Aquila,Italy}

\begin{abstract}
Although ferroelectric compounds containing hydrogen bonds were
among the first to be discovered, organic ferroelectrics are relatively
rare.  The discovery of high polarization at room temperature in
croconic acid [Nature \textbf{463}, 789 (2010)] has led to a renewed
interest in organic ferroelectrics. We present an ab-initio study
of two ferroelectric organic molecular crystals,
1-cyclobutene-1,2-dicarboxylic acid (CBDC) and 2-phenylmalondialdehyde
(PhMDA). By using a distortion-mode analysis we shed light on the
microscopic mechanisms contributing to the polarization, which we
find to be as large as 14.3 and 7.0\,$\mu$C/cm$^{2}$ for CBDC
and PhMDA respectively. These results suggest that it may be
fruitful to search among known but poorly characterized organic
compounds for organic ferroelectrics with enhanced polar properties
suitable for device applications.
\end{abstract}

\date{\today}


\maketitle


The property of polarization switchable by an applied external
electric field, i.e., ferroelectricity, is the basis
of a wide range of device applications\cite{ref1}. The first known ferroelectric material, discovered
in 1920, was sodium potassium tartrate tetrahydrate
(NaKC$_{4}$H$_{4}$O$_{6}\cdot$H$_{2}$O), better known as Rochelle
salt\cite{ref2}. Unfortunately it is also one of the most
complicated ferroelectrics known to date, and research in
this field soon focused on simpler ferroelectrics, such as phosphates
and arsenates, discovered later\cite{ref4,ref5}. A prototypical
example is potassium dihydrogen phosphate, KH$_{2}$PO$_{4}$,
better known as KDP\cite{Tosatti}. The latter contains hydrogen
bonds for which different possible arrangements of the hydrogens
can result in different orientations of the dipolar units. After
the discovery of ferroelectricity in BaTiO$_{3}$, with polarizations
as large as 27\,$\mu$C/cm$^{2}$ in the tetragonal phase,\cite{BTO}
attention switched to this new class of perovskite oxygen octahedral
ferroelectrics made up from basic BO$_{6}$ building blocks, of which
BaTiO$_3$ was the forerunner.  These perovskite and related materials are
by far the most investigated class of ferroelectrics, and the most
important for current device applications.  Nevertheless, substantial
efforts are now being made in order to find ferroelectric materials
that are potentially cheaper, more soluble, less toxic, lighter, or
more flexible\cite{ref6,ref7}.
Very recently it was discovered that organic crystals of
croconic acid, H$_{2}$C$_{5}$O$_{5}$, exhibit ferroelectricitly
with a large spontaneous polarization of 21\,$\mu$C/cm$^{2}$
\cite{Horiuchi1,Horiuchi2}. Croconic acid consists of polar
stacks of sheets of hydrogen-bonded molecules\cite{Braga}.
Upon application of an electric field, protons associated
with one molecule cooperatively shift to a hydrogen-bonded
neighbor, switching the molecular dipole and giving rise to a
large polarization\cite{Horiuchi1}. It seems likely that this
physical mechanism may also occur in many other organic materials
whose ferroelectric properties have yet to be discovered and
characterized. Ab-initio calculations are a suitable tool in
this case, as they can not only identify materials with large
polarization, but also shed light on the mechanisms behind the
polarization itself.

In this work, we focus on 1-cyclobutene-1,2-dicarboxylic acid
(CBDC, C$_{6}$H$_{6}$O$_{4}$)\cite{CBDC2} and 2-phenylmalondialdehyde
(PhMDA, C$_{9}$H$_{8}$O$_{2}$)\cite{PhMDA1}.
Initial x-ray diffraction studies
of CBDC\cite{CBDC2} and PhMDA\cite{PhMDA1} were performed only
at room temperature for CBDC and at $T$=111\,K for PhMDA. It was only
recently that new measurements were extended to low temperature
for CBDC and to room temperature for PhMDA,\cite{AdvancedMat}
showing that both compounds are \textit{ferroelectric} with Curie
temperatures above 400 and 363\,K respectively. The observed spontaneous
polarization ($P_{\textrm{x}}$,$P_{\textrm{y}}$,$P_{\textrm{z}}$)
is as large as (0.4,0,2.8) and (0,0,9)\,$\mu$C/cm$^{2}$, for CBDC
and PhMDA respectively\cite{AdvancedMat}.
The main purpose of the present study is to describe the ferroelectricity
in these compounds based on ab-initio calculations.
We have found that for both materials, the main
contribution to the polarization is a collective proton transfer
between molecular units, hereafter denoted as an ``inter-molecular''
hydrogen shift. This is basically the same main mechanism found in
croconic acid. However, importantly, we have also introduced a partial
mode analysis of the contributions to the ferroelectricity,
and found that correlated intramolecular distortions also contribute
significantly to the polarization.

First-principles density functional theory calculations were performed
using the Vienna-Ab-Initio Simulation package (VASP)\cite{vasp1,paw}. The
Kohn-Sham equations were solved using the projector augmented wave
pseudo-potentials and the PBE generalized gradient density approximation
to the exchange-correlation potential\cite{perdew}. We used
a plane-wave cut-off of 400\,eV and k-point meshes of (6,2,4) and
(4,2,6) for CBDC and PhMDA respectively. The Berry-phase method was
employed to evaluate the crystalline polarization\cite{Vanderbilt,Resta1}.
All atomic
positions were optimized until the forces were below 0.01\,eV/\AA.
Test calculations were performed to estimate the effect of (1)
effects beyond the local density approximation by
using the Heyd-Scuseria-Ernzerhof hybrid functional as implemented
in VASP,\cite{hse2,hse3,hse5} and (2)
van der Waals corrections\cite{vdw1,vdw2} as proposed within a
density-functional framework. In both cases the
changes in the magnitude of polarization were found to be less than
a few percent, confirming the basic physics explained in the present
work.
\begin{figure}
\centerline{\includegraphics[width=1.10\columnwidth,angle=0]{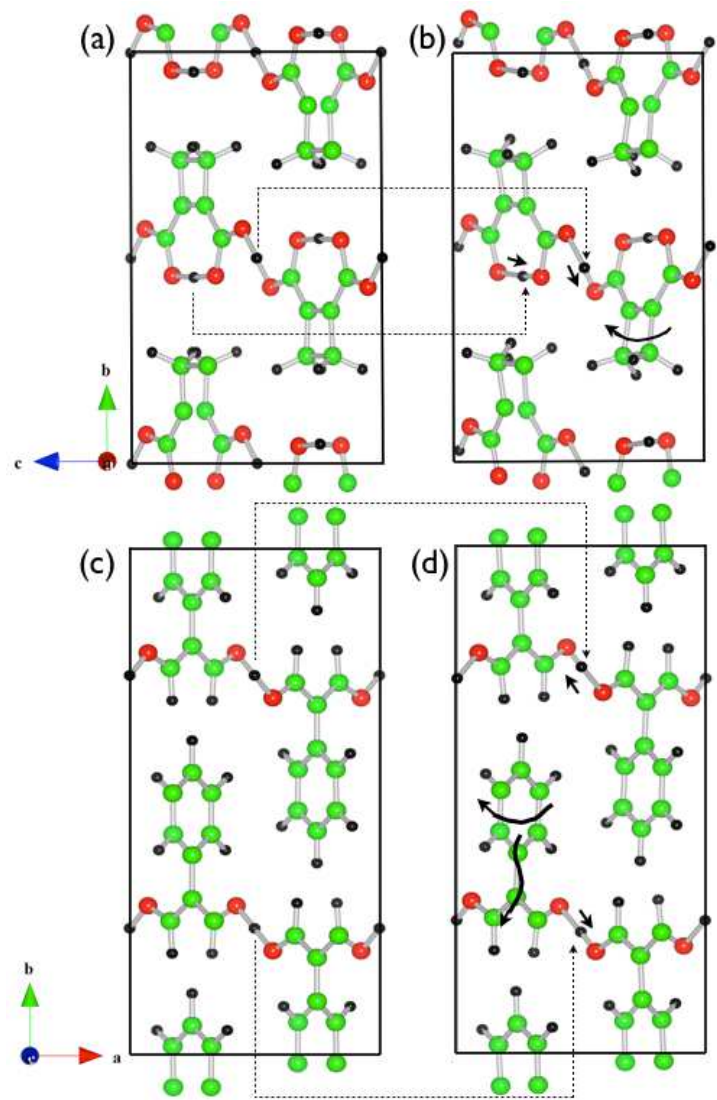}}
\caption{(Color online) Ball-and-stick models of CBDC (top) and
PhMDA (bottom).  (a) and (c) refer to centric and (b) and (d) to
polar structures.  Dashed guiding-eye lines refer to the position
of relevant hydrogens contributing to polarization; arrows in
(b) and (d) indicate important polar distortions.}\label{fig1}
\end{figure}

In Fig.~\ref{fig1} we show a ball-and-stick model of CBDC (top)
and PhMDA (bottom). In (a) and (c) we show the centric structures,
and in (b) and (d) the polar ones. The CBDC molecular unit is
formed by a main planar four-membered ring of carbon atoms similar
to the planar cyclobutene molecule, and two carboxyl carbons,
i.e., with formula $-$C$=$O$-$O$-$H, where `$=$' and
`$-$' refer to double and single bonds. The molecular units
shown in Fig.~\ref{fig1} are part of an infinite chain of molecules
related by a glide plane and linked by intermolecular hydrogen
bonds. The PhMDA molecular unit is formed by a planar phenyl group,
i.e., six carbon atoms arranged in a planar ring, each of
which is bonded to an H atom. The phenyl group is linked to a
linear hydrogen-bonded chain of $\beta$-diketone enol moieties,
i.e., O($=$)C$-$C$=$C$-$O$-$H. Hydrogen bonds link the molecules
together along the [102] and [10$\overline{2}$] directions into
infinite chains.

We found it useful here to introduce a pseudosymmetry analysis,
in which a given low-symmetry (ferroelectric) structure is represented in terms
of a symmetry-lowering Landau-type structural phase transition
from a high-symmetry (paraelectric) parent structure, based on finding a supergroup
of the given space group\cite{bilbao4,explan-pseudosymm}. 
CBDC crystallizes in the
monoclinic space group $Cc$ and its pseudosymmetric centric
structure has space group symmetry $C2/c$; the two structures are
related by a maximum atomic distortion of 0.43\,\AA. Analogously,
PhMDA crystallizes in the $Pna21$ and its pseudosymmetric centric
structure is $Pbcn$, with maximum atomic displacements of 0.25\,\AA.
In order to gain insight into the ferroelectricity, we compare the
relaxed centric and polar structures shown in Fig.~\ref{fig1}(a-b)
for CBDC and in Fig.~\ref{fig1}(c-d) for PhMDA.  For CBDC
we have two types of hydrogen bonds (``intramolecular'' and
``intermolecular'') which, in the polar state, shift toward the
molecular units on the right, as shown by the short arrows
in Fig.~\ref{fig1}(b). There is another
cooperative atomic distortion, shown in Fig.~\ref{fig1}(b) by the
curved arrow, hereafter referred to ``molecular buckling.'' For
PhMDA, the hydrogen sitting between two molecular units
in the polar structure shifts towards one of its neighboring units,
as indicated by short arrows in Fig.~\ref{fig1}(d).  Two other
relevant atomic distortions come into play, as shown by the curved
arrows, both tending to deform the molecular units.  One acts on
the planar phenyl group, while the other distorts the
$\beta$-diketone enol moieties. For both compounds, then, we find
that three different types of distortion can contribute
importantly to the ferroelectric polarization.

\begin{figure}
\centerline{\includegraphics[width=1.0\columnwidth,angle=0]{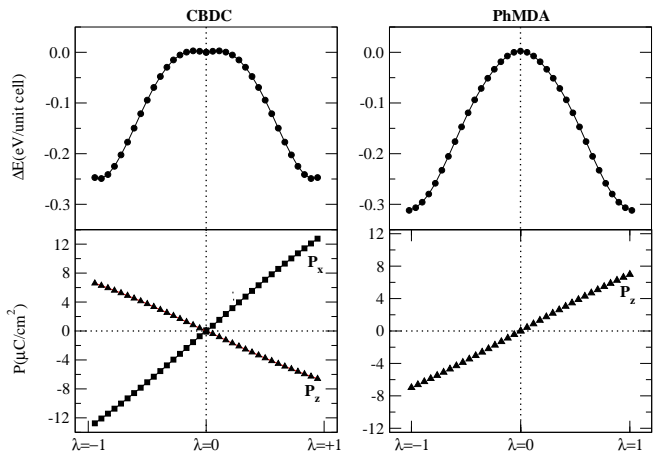}}
\caption{Variation of total energy (top) and of polarization
(bottom) as a function of the amplitude of the polar distortion
between centric ($\lambda$=0) and polar ($\lambda$=$\pm$1)
configurations.}\label{fig2}
\end{figure}

In Fig.~\ref{fig2} we show the variation of the total energy from the
centric to the polar structure as a function of the amplitude of
the polar mode. For both materials we find a bi-stable energy profile
characteristic of a ferroelectric material with an energy barrier
on the order of 0.3\,eV/unit cell, suggesting that the polarization
should be
switchable upon application of an external electric field. For
CBDC, the polarization is in the $ac$ plane with a magnitude of
$P$=14.3\,$\mu$C/cm$^{2}$, while for PhMDA it is along the $c$ axis
and equal to 7.0\,$\mu$C/cm$^{2}$. (Recall that the polarization
of croconic acid is 21\,$\mu$C/cm$^{2}$).  The slight discrepancy with 
experimental values on
CBDC might be due to insufficient optimization of high-quality-
crystallization.cite{AdvancedMat}. 
We next consider the relaxed structures of the high- and low-symmetry
phases, analyzing the displacive-type transition between the two
phases in terms of symmetry modes using the Amplimodes
software package\cite{bilbao7}.
The program determines the global structural distortion that relates
the two phases, enumerates the symmetry modes compatible with the
symmetry breaking, and decomposes the total distortion into amplitudes
of these orthonormal symmetry modes.

\begin{figure}
\centerline{\includegraphics[width=1.0\columnwidth,angle=0]
{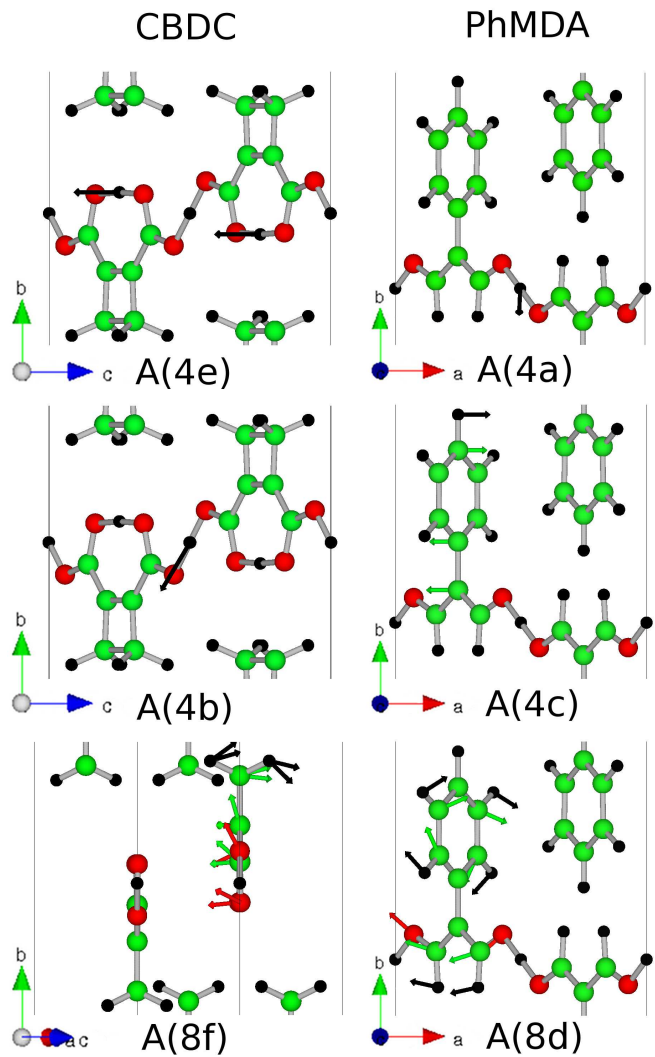}}
\caption{(Color online) Displacements patterns (arrows)
connecting centric to polar structures for atoms belonging to
specified Wyckoff positions (top to bottom) for CBDC (left) and PhMDA
(right).}\label{fig3}
\end{figure}

\begin{figure}
\centerline{\includegraphics[width=.8\columnwidth,angle=-90]
{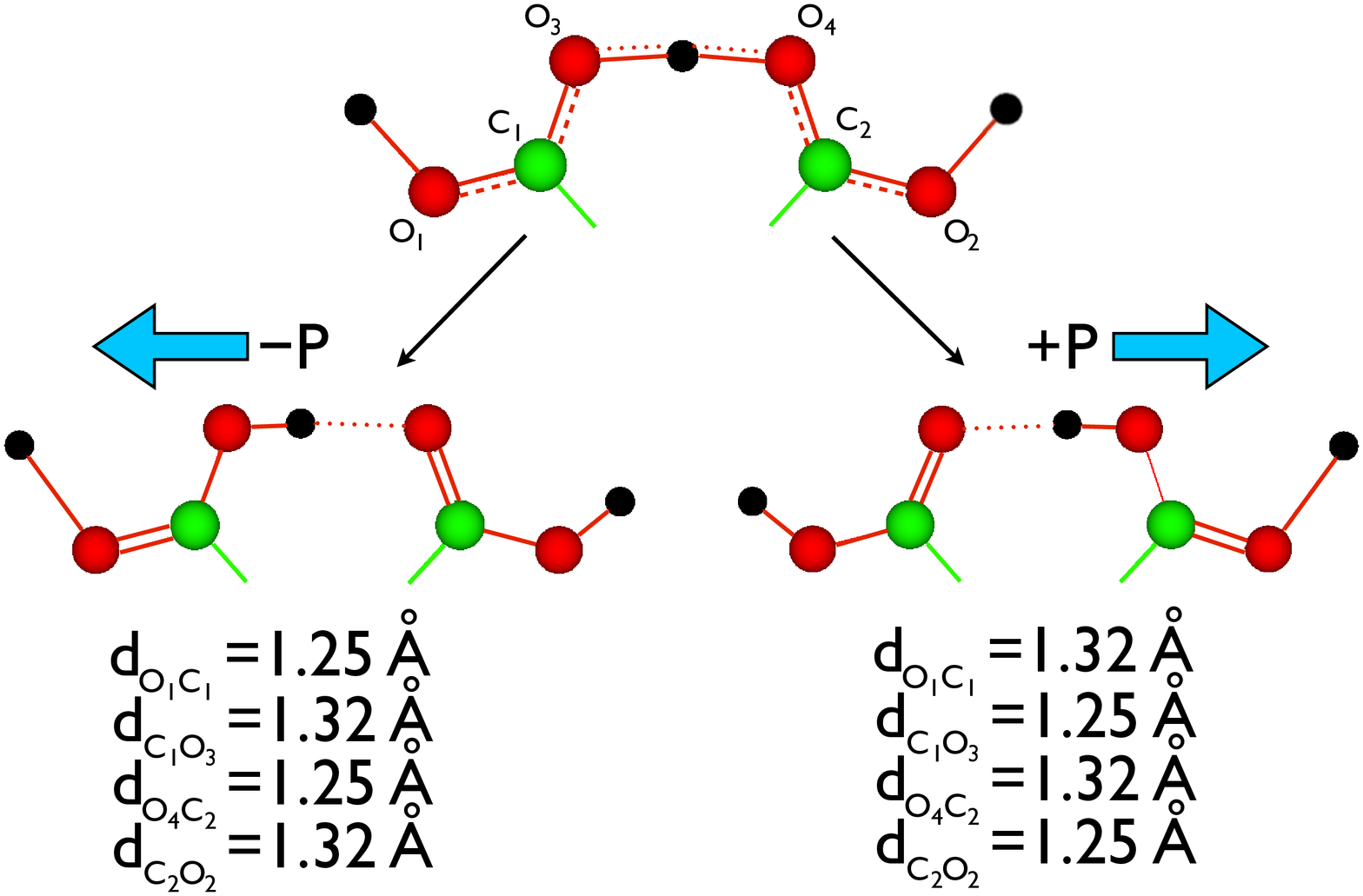}}
\caption{(Color online) Switching between double and single bonds 
in the CBDC molecule. Top part: centric case; bottom part: switching between $+P$ and $-P$. 
The chemist convention for the orientation of the dipole moment is used here, \textit{i.e.} arrow starts at $\delta+$ and ends at $\delta-$.}\label{fig3.5}
\end{figure}

In Fig.~\ref{fig3} we show the centric structures for CBDC (left)
and PhMDA (right), with the characteristic atomic displacements of
the different polar distortion modes shown by colored arrows.
As the polar mode acts separately on different Wyckoff positions
(WPs) of the high-symmetry structure, it is meaningful to consider
the action of the polar distortion on atoms belonging to different
WPs separately. We denote these as $A$(WP) and they are shown from
top to bottom in Fig.~\ref{fig3}.  For CBDC, we have $A(4e)$ and
$A(4b)$ as intra- and inter-molecular proton transfer distortions and $A(8f)$
as out-of-plane molecular twisting (buckling) and pi-bond
switching of carboxylic groups. For each of them we have calculated the
polarization by displacing \textit{only} the atoms belonging to a
given WP orbit and keeping the rest of them in their centrosymmetric
positions, obtaining
${\bf P}_{\textrm{4b}}$=(6.6,0,$-$5.5),
${\bf P}_{\textrm{4e}}$=(0.5,0,$-$1.6) and
${\bf P}_{\textrm{8f}}$=(5.5,0,0.4)\,$\mu$C/cm$^{2}$. Notably, their sum
is (12.6,0,$-$6.7)\,$\mu$C/cm$^{2}$ which is almost equal to the total
polarization ${\bf P}_{\textrm{tot}}$=(12.7,0,$-$6.6)\,$\mu$C/cm$^{2}$
calculated from the total polar distortion.
The linearity of the partial polarizations is compatible with
displacive-type ferroelectricity.
Surprisingly, the mode decomposition
shows that the A (8f) mode makes almost as large a contribution
as the inter-molecular proton transfer in determining the total polarization.
This effect can be related to a double($\pi$)-bond
 switching of carboxylic C=O$\Longleftrightarrow$C$-$O bonds correlated 
with the ``inter-molecular'' \textit{and} ``intra-molecular'' hydrogen distortion.
This is shown in 
Fig.\ \ref{fig3.5}, where we present an enlarged view of the
$A({\textrm{8f}})$ mode. In the upper part, 
we show the centric case, with hydrogens equidistant from nearest carbons or oxygens.
In the lower part, we show the cooperative hydrogen distortions leading to the $+{\bf P}$ or $-{\bf P}$ state,
 which correlates, in turn, with  the switching of double and single C-O bonds
and with the shortening/elongation 
of the corresponding C-O bonds, in both  $+{\bf P}$ and $-{\bf P}$.
This picture is further confirmed 
by the following computational experiment: (i) we first consider 
 all the atoms at their centric positions (upper part in Fig.\ \ref{fig3.5}); 
(ii) we then move only the intra-molecular hydrogen as, for instance, in
the $+{\bf P}$ state, keeping  all the other atoms fixed.
The charge density  difference between (i) and (ii) (not reported here) 
shows an incipient pile-up of out-of-plane charge between C$_2$=O$_2$ and 
C$_1$=O$_3$, which corresponds to the incipient formation of 
the $\pi$ (double) bonds. 
One could expect that the  polar carboxylic groups, rather than the less polar C-C
 bonds, might be responsible for  the large polarization of the buckling mode. 
To confirm this,
we have further extracted the contribution of the C$=$O/C$-$O
bonds to this polarization.
   The extracted polarization of (4.8, 0, -0.1)$\mu$C/cm$^2$ [vs. the
remaining contributions of
   (0.8, 0, 0.4)\,$\mu$C/cm$^{2}$] account for most of the large
polarization of the A(8f) mode.
Again linearity is well fulfilled. This clearly explains the origin of the 
surprisingly large polarization of the buckling mode.
For PhMDA, we find partial modes $A(4a)$, $A(4c)$ and $A(8d)$
whose contributions to the polarization are 5.8, 1.0 and
0.3\,$\mu$C/cm$^{2}$ along the $z$ polar axis, respectively. The
linearity holds also in this case, but now the inter-molecular
proton transfer does give the dominant contribution.

Finally, for the centric structures, we have calculated the dimensionless
Born (or ``dynamical'' or ``infrared'') charge tensors
$Z_{k,\alpha\beta}^{*}=(\Omega/e)\partial P_{\alpha}/\partial
u_{k,\beta}$, where $P_{\alpha}$ is the component of the polarization
in direction $\alpha$ and $u_{k,\beta}$ is the displacement of
atom $k$ in direction $\beta$, $\Omega$ is the primitive cell
volume, and $e$ is the charge quantum.  
In perovskite ABO$_{3}$ oxides the ferroelectric tendency is well
known to be connected with the presence of anomalously large Born
charges\cite{Resta1,Vanderbilt}.
It should be noted that in low-symmetry cases, as in the present
study, the Born tensor is not symmetric in its Cartesian indices.
Therefore, we have split the tensor into symmetric $Z^{*}_{S}$
and antisymmetric $Z^{*}_{AS}$ parts.  In the following, we will
focus on the former, and in particular, on its three eigenvalues
$\lambda_1>\lambda_2>\lambda_3$ whose physical interpretation is
straightforward. Furthermore, only the hydrogen Born tensors will be considered.  
We have also calculated the phonon frequencies at
the $\Gamma$ point; the presence of an imaginary frequency usually
implies a structural instability, in this case of the paraelectric
structure.

Let us first consider CBDC. As expected, the significant
deviations of the dynamical tensor with respect to the nominal
charges involve the ``active'' H atoms. For the ``inter-molecular''
hydrogen, $\lambda_{i,{Z^{*}_{S}}}$=(3.4,0.4,0.1) and for the
``intra-molecular'' hydrogen, $\lambda_{i,{Z^{*}_{S}}}$=(2.2,0.3,0.3).
The large values of the Born charges
for hydrogens confirm their important contribution to the polarization, as found also 
in other hydrogen-bonded organic
molecular crystals\cite{Colizzi1,Ishii}.
The other hydrogens have only negligible absolute eigenvalues $\sim$ 0.1  
(some of them becoming  negative).
For the phonons, we found a large non-degenerate imaginary $\Gamma$
phonon frequency of about 106\,cm$^{-1}$. According to a symmetry
analysis,\cite{sam} infrared irreducible representations exist for
\textit{all} three WP positions with either $A_{u}$ or $B_{u}$
symmetry. This is not unexpected, as all WP positions carry a
contribution to the polarization. In particular, the eigenvector
of the imaginary frequency has symmetry $B_{u}$, which is
\textit{polar}. After normalization to 1\,\AA, we use the Amplimode
software for studying the corresponding displacement pattern. The
largest absolute $|u|$, where $u$ is the displacement of the atom
according to the phonon eigenvector, is 0.29 and 0.35\,\AA\ for
``intra-molecular'' and `` inter-molecular'' hydrogens. Again, this
confirms the dominant role of the two types of hydrogen in the
ferroelectric properties.
For the case of PhMDA, we found significant deviations of the dynamical
charge tensor for ``inter-molecular'' hydrogen, whose eigenvalues are
$\lambda_{i,{Z^{*}_{S}}}$=(4.1, 0.4, 0.2). The eigenvalues for 
other hydrogen atoms are  smaller than 1. Finally,  
the imaginary phonon frequency is 112\,cm$^{-1}$
with \textit{polar} symmetry $B_{u}$. Also in this case,
the polarization vector of the eigenmode has a large displacement
of $\sim$\,0.43\,\AA\ for the ``intermolecular'' hydrogen.

To summarize, we have studied the origin of ferroelectricity in
CBDC and PhMDA using first-principles calculations and symmetry
analysis methods. The estimated polarizations are as large as 
$\sim$\,14\,$\mu$C/cm$^{2}$ for CBDC and $\sim$\,7\,$\mu$C/cm$^{2}$ for
PhMDA.  We have shown that a partial mode analysis is a useful tool
for exploring the polarization mechanisms. In both compounds, the
proton transfer between (or within) molecular units appears to be
the main contribution, as confirmed by the large dynamical charges
and the analysis of the eigenmode displacement patterns. However,
other contributions, especially the pi-bond switching of carboxylic groups in
CBDC may also
have a significant weight in the final polarization.  Again, partial mode analysis 
has been used to elucidate the origin of the unexpectedly large contribution. 
 We hope that our study will stimulate further attempts to search for
new organic ferroelectrics with potentially large electric
polarizations.

The work is supported by the European Research Council, 7th Framework
Programme - FP7 (2007-2013)/ERC Grant Agreement n. 203523 and 
partly supported by the FIRST program from JSPS.
Computational
support by CASPUR Supercomputing Center in Rome is gratefully
acknowledged. D.V. acknowledges support
of ONR Grant N-00014-05-1-0054. S.H. acknowledges the support by KAKENHI (20110003) and the Sumitomo Foundation.
A.S. thanks J.M. Perez-Mato, E. Tasci, D. Orobengoa, H. Stokes\cite{Stokes} and I.
Baburin for useful discussions about symmetry aspects of the study.

\bibliography{biblio}

\end{document}